\documentclass[]{pasj01}
\draft
\begin{document} 
\Received{}
\Accepted{}
\title{Scale heights and equivalent widths of the iron K-shell lines in the Galactic diffuse X-ray emission}

\author{
Shigeo \textsc{Yamauchi}\altaffilmark{1$\ast$},
Kumiko K. \textsc{Nobukawa}\altaffilmark{2},
Masayoshi \textsc{Nobukawa}\altaffilmark{3},
Hideki \textsc{Uchiyama}\altaffilmark{4},
and Katsuji \textsc{Koyama}\altaffilmark{2, 5}
}
\altaffiltext{1}{Department of Physics, Nara Women's University, Kitauoyanishimachi, Nara 630-8506}
\email{yamauchi@cc.nara-wu.ac.jp}
\altaffiltext{2}{Department of Physics, Graduate School of Science, Kyoto University, \\
Kitashirakawa-oiwake-cho, Sakyo-ku, Kyoto 606-8502}
\altaffiltext{3}{Department of Teacher Training and School Education, Nara University of Education, Takabatake-cho, Nara 630-8528}
\altaffiltext{4}{Faculty of Education, Shizuoka University, 836 Ohya, Suruga-ku, Shizuoka 422-8529}
\altaffiltext{5}{Department of Earth and Space Science, Graduate School of Science, Osaka University, \\
1-1 Machikaneyama-cho, Toyonaka, Osaka 560-0043}
\maketitle

\begin{abstract}

This paper reports the analysis of the X-ray spectra of the Galactic diffuse X-ray emission (GDXE) in the Suzaku archive. 
The fluxes of the Fe\,\emissiontype{I}\,K$\alpha$ (6.4 keV),
 Fe\,\emissiontype{XXV}\,He$\alpha$ (6.7 keV)
and Fe\,\emissiontype{XXVI}\,Ly$\alpha$ (6.97 keV) lines are separately determined. 
From the latitude distributions, we confirm that  the GDXE is decomposed into the Galactic center (GCXE), 
the Galactic bulge (GBXE) and  the Galactic ridge (GRXE) X-ray emissions.
The scale heights (SHs) of the Fe\,\emissiontype{XXV}\,He$\alpha$ line of the GCXE, GBXE and GRXE are 
determined to be $\sim$40, $\sim$310 and $\sim$140~pc, 
while those of the Fe\,\emissiontype{I}\,K$\alpha$ line are $\sim$30, $\sim$160 and $\sim$70~pc, respectively.
The mean equivalent widths (EWs)  of the sum of the
Fe\,\emissiontype{XXV}\,He$\alpha$ and Fe\,\emissiontype{XXVI}\,Ly$\alpha$ lines
are $\sim$750~eV, $\sim$600~eV and $\sim$550~eV, while those of the
Fe\,\emissiontype{I}\,K$\alpha$ line are
$\sim$150~eV, $\sim$60~eV and $\sim$100~eV for the GCXE, GBXE and GRXE, respectively.
The origin of the GBXE, GRXE and GCXE is separately discussed based on  the new results of the SHs and EWs, 
in comparison with those of the Cataclysmic Variables (CVs), 
Active Binaries (ABs) and Coronal Active stars (CAs).
\end{abstract}

\section{Introduction}

The Galactic diffuse X-ray emission (GDXE) is unresolved X-ray emission along the Galactic plane \citep{Worrall1982,Warwick1985}. 
Strong K-shell lines from highly ionized atoms were found in the GDXE spectra
\citep{Koyama1986,Koyama1989,Yamauchi1990,Yamauchi1993,Kaneda1997,Sugizaki2001,
Ebisawa2005}. Subsequently, the K-shell lines were resolved into Helium-like and Hydrogen-like atomic lines, such as 
Fe\,\emissiontype{XXV}\,He$\alpha$, 
Fe\,\emissiontype{XXVI}\,Ly$\alpha$, 
S~\emissiontype{XV}\,He$\alpha$ and 
S~\emissiontype{XVI}\,Ly$\alpha$ 
\citep{Koyama1996,Koyama2007b,Ebisawa2008, Heard2013a}.  
The Fe\,\emissiontype{XXV}\,He$\alpha$ (6.7~keV) 
and Fe\,\emissiontype{XXVI}\,Ly$\alpha$ (6.97 keV) lines are 
emitted from a high temperature plasma (HP) of $kT\sim$5--7~keV, while the S~\emissiontype{XV}\,He$\alpha$ (2.46 keV) 
and S~\emissiontype{XVI}\,Ly$\alpha$ (2.62 keV) lines 
come from a low temperature plasma (LP) of $kT\sim$1~keV. 
\citet{Koyama1996} discovered  Fe\,\emissiontype{I}\,K$\alpha$ (6.4 keV) lines from the Galactic center (GC) region.  
Bright regions are associated with molecular clouds, 
hence the origin is fluorescence from cool gas (CG). 
They are called as the X-ray reflection nebulae (XRNe). 
Furthermore, \citet{Ebisawa2008} and \citet{Yamauchi2009} found the Fe\,\emissiontype{I}\,K$\alpha$ line in the various regions 
along the Galactic plane. 
Therefore the Fe\,\emissiontype{I}\,K$\alpha$ line emission is not only from XRNe, 
but its large fraction is more extended emission along  Galactic plane.
From the spatial distributions of the Fe K-shell lines, \citet{Koyama1989}, \citet{Yamauchi1993} and \citet{Uchiyama2013} 
decomposed the GDXE into the Galactic Center (GCXE), Galactic Bulge (GBXE) and  Galactic Ridge (GRXE) X-ray Emissions 

Long standing debates have been the origin of the GDXE. 
Most of the previous debates were based on the observations of limited spatial and spectral resolution, 
where  the GDXE was not separated into the GRXE, GBXE and GCXE.  
The K-shell line emission at $\sim$6.7~keV was not resolved to the Fe\,\emissiontype{I}\,K$\alpha$, Fe\,\emissiontype{XXV}\,He$\alpha$ 
and Fe\,\emissiontype{XXVI}\,Ly$\alpha$ lines.

In this paper, we analyze the Suzaku archive data from a large number of pointing positions  along the inner Galactic plane. 
We confirm that the GDXE is composed of the GCXE, GBXE and GRXE. 
Furthermore we separately determine the scale heights (SHs) and equivalent widths (EWs) of the Fe\,\emissiontype{I}\,K$\alpha$,
Fe\,\emissiontype{XXV}\,He$\alpha$ and 
Fe\,\emissiontype{XXVI}\,Ly$\alpha$ lines in the GCXE, GBXE and GRXE.
Based on these results, we examine the origin of the HP and CG in the GCXE, GBXE and GRXE.  
Throughout this paper, the distance to the GC is 8 kpc and quoted errors are in the 68\% (1$\sigma$) confidence limits.   
	
\section{Observations and Data Reductions}

\begin{longtable}{*{4}{c}}
\caption{Observation logs.}
\hline 
Sequence No. & Pointing position & Observation time (UT)& Exposure \\ 
                       &  $l$, $b$  &  Start -- End                & (ksec)   \\ 
\hline
\endfirsthead
\hline
Sequence No. & Pointing position & Observation time (UT)& Exposure \\ 
                       &  $l$, $b$  &  Start -- End                & (ksec)     \\ 
\hline
\endhead
\hline
\endfoot
\hline
\endlastfoot
407094010& 	322.06,	$-$0.42	&2012-08-15 15:34:02 --  2012-08-16 00:11:17& 	30.0 		\\
501043010&	330.40, 	$-$0.38 	&2006-09-16 11:02:03 --	2006-09-17 07:14:14&	43.6 		\\
503073010& 	331.30,	$-$0.76 	&2008-09-20 18:18:35 --  2008-09-21 13:30:14& 	53.7 		\\
503074010& 	331.47,	$-$0.64 	&2008-09-21 13:31:03 --  2008-09-22 06:40:12& 	52.6 		\\
501042010& 	331.57,	$-$0.53 	&2006-09-15 16:00:48 --	2006-09-16 10:58:14& 	40.2 		\\
100028020&	332.00, 	$-$0.15 	&2005-09-18 22:47:36 --	2005-09-19 11:58:41&	19.3 		\\
100028010&	332.40, 	$-$0.15 	&2005-09-19 12:00:02 --	2005-09-20 19:38:24&	41.4 		\\
100028030&	332.70, 	$-$0.15 	&2005-09-20 19:40:17 --	2005-09-21 07:29:24&	21.9 		\\
401056010&	333.54, 	0.33 		&2006-09-20 20:25:12 --	2006-09-21 17:21:20&	39.1 		\\
407018010& 	333.61,	$-$0.20 	&2012-08-21 23:55:27 --  2012-08-22 22:33:20& 	40.5 		\\
407020010& 	333.72,	0.22 		&2012-08-19 12:30:06 --  2012-08-20 12:35:23& 	44.3 		\\
407091010& 	333.89,	0.41 		&2012-08-18 19:16:15 --  2012-08-19 12:29:17& 	29.3 		\\
507068010& 	337.21,	$-$0.73 	&2012-09-02 13:16:23 --  2012-09-11 06:14:09& 	304.2 	\\
404056010&	338.00, 	0.08 		&2010-03-12 23:40:40 --	2010-03-14 10:49:14&	50.6 		\\
505049010& 	339.01,	$-$0.93 	&2010-09-10 18:33:35 --  2010-09-12 04:50:16& 	51.9 		\\
505051010& 	339.43,	$-$0.80 	&2010-09-23 06:09:00 --  2010-09-24 09:28:16& 	50.2 		\\
505050010& 	339.79,	$-$1.14 	&2010-09-12 04:52:53 --  2010-09-13 14:35:12& 	52.7 		\\
401052010&	340.05, 	0.13 		&2006-09-09 09:12:56 --	2006-09-09 22:05:14&	22.5 		\\
406078010& 	340.17,	$-$0.12 	&2012-02-23 22:39:06 --  2012-02-26 12:52:21& 	149.8 	\\
405027010&	340.44, 	$-$0.18 	&2011-02-11 03:42:23 --	2011-02-11 18:16:19&	20.9 		\\
505052010& 	340.77,	$-$1.01 	&2010-09-24 09:32:21 --  2010-09-25 12:29:19& 	49.6 		\\
401054010&	341.37, 	0.60 		&2006-10-05 21:10:30 --	2006-10-06 10:05:24&	21.1 		\\
502049010&	344.26, 	$-$0.22 	&2008-03-25 11:00:23 --	2008-03-30 15:00:14&	215.7 	\\
100026030&	345.80, 	$-$0.54 	&2005-09-28 07:09:13 --	2005-09-29 04:25:24&	37.5 		\\
100026020&	347.63, 	0.71 		&2005-09-25 19:11:40 --	2005-09-26 15:42:09&	34.9 		\\
505076010& 	347.85,	$-$0.23 	&2011-02-16 01:17:29 --  2011-02-16 23:10:11& 	32.6 		\\
501105010& 	348.80,	$-$0.54 	&2007-02-23 08:36:38 --  2007-02-23 18:39:24& 	20.7 		\\
503108010&	348.92, 	$-$0.45 	&2008-08-28 05:36:23 --	2008-08-28 22:17:14&	23.5 		\\
408021010&	352.17,	$-$0.27 	&2013-09-05 00:04:43 --  2013-09-05 19:14:14&	37.3 		\\
405026010&	355.27, 	0.39 		&2011-02-19 23:54:57 --	2011-02-20 16:43:24&	20.9 		\\
503022010&	356.00, 	0.70 		&2009-03-18 23:11:24 --	2009-03-19 21:26:24&	41.3 		\\
504049010&	356.30, 	1.00 		&2009-09-08 03:36:55 --	2009-09-09 02:36:11&	18.2 		\\
503023010&	356.33, 	0.70 		&2009-03-26 06:37:01 --	2009-03-27 01:53:19&	31.2 		\\
503020010&	356.40, 	$-$0.05 	&2009-02-21 01:15:55 --	2009-02-22 18:59:14&	61.1 		\\
505082010&	356.40, 	$-$0.40 	&2011-03-15 13:54:28 --	2011-03-16 14:54:23&	48.5 		\\
505083010&	356.40, 	$-$0.80 	&2010-10-10 14:04:38 --	2010-10-11 21:33:13&	52.9 		\\
505084010&	356.40, 	$-$1.50 	&2011-03-06 05:36:44 --	2011-03-07 13:01:11&	50.3 		\\
505085010&	356.40, 	$-$2.30 	&2010-10-13 06:04:47 --	2010-10-14 13:30:17&	55.0 		\\
505086010&	356.40, 	$-$3.50 	&2010-10-14 13:31:50 --	2010-10-16 01:30:25&	54.1 		\\
503019010&	356.65, 	$-$0.05 	&2009-02-19 16:37:49 --	2009-02-21 01:15:14&	52.8 		\\
503018010&	356.90, 	$-$0.05 	&2008-09-24 09:27:54 --	2008-09-24 22:30:24&	29.4 		\\
503018020&	356.90, 	$-$0.05	&2008-10-03 18:05:13 --	2008-10-04 03:42:18&	12.2		\\
503018030&	356.90, 	$-$0.05	&2009-02-19 07:32:01 --	2009-02-19 16:36:24&	11.9		\\
503017010&	357.15, 	$-$0.05 	&2008-09-23 08:08:10 --	2008-09-24 09:21:13&	51.3 		\\
503016010&	357.40, 	$-$0.05 	&2008-09-22 06:47:49 --	2008-09-23 08:07:17&	52.2 		\\
503015010&	357.65, 	$-$0.05 	&2008-09-19 07:33:05 --	2008-09-20 09:56:13&	56.8 		\\
504036010& 	357.71,	$-$0.12 	&2009-08-29 12:05:20 --  2009-09-01 00:13:24& 	136.5 	\\
503014010&	357.90, 	$-$0.05 	&2008-09-18 04:46:49 --	2008-09-19 07:32:20&	55.4 		\\
501053010&	358.17, 	0.00 		&2006-10-10 21:18:59 --	2006-10-11 10:06:14&	21.9 		\\
504002010&	358.47, 	$-$0.59 	&2010-02-27 16:14:41 --	2010-02-28 22:50:14&	53.1 		\\
501052010&	358.50, 	0.00 		&2006-10-10 06:45:09 --	2006-10-10 21:18:14&	19.3 		\\
504090010&	358.50, 	$-$1.20 	&2009-10-13 04:17:20 --	2009-10-14 11:29:06&	52.9 		\\
504091010&	358.50, 	$-$1.60 	&2009-09-14 19:37:36 --	2009-09-16 07:18:14&	51.3 		\\
504092010&	358.50, 	$-$2.15 	&2009-09-16 07:21:35 --	2009-09-17 13:49:14&	50.9 		\\
504093010&	358.50, 	$-$2.80 	&2009-09-17 13:54:31 --	2009-09-19 03:37:14&	53.2 		\\
504094010&	358.50, 	$-$3.80 	&2009-09-19 03:40:31 --	2009-09-21 14:33:19&	93.1 		\\
504095010&	358.50, 	$-$5.00 	&2009-10-15 15:32:28 --	2009-10-16 18:00:19&	48.3 		\\
504001010&	358.53, 	$-$0.27 	&2010-02-26 09:15:00 --	2010-02-27 16:13:16&	51.2 		\\
504003010&	358.55, 	$-$0.87 	&2010-02-25 04:33:17 --	2010-02-26 09:13:19&	50.9 		\\
501051010&	358.83, 	0.00 		&2006-10-09 13:40:09 --	2006-10-10 06:44:24&	21.9 		\\
500019010&	358.91, 	$-$0.04 	&2006-02-23 10:51:11 --	2006-02-23 20:02:19&	13.3 		\\
500018010&	359.43, 	$-$0.09 	&2006-02-20 12:45:25 --	2006-02-23 10:50:14&	106.9 	\\
503072010&	359.58, 	0.17 		&2009-03-06 02:39:12 --	2009-03-09 02:55:25&	140.6 	\\
100027020&	359.75, 	$-$0.05 	&2005-09-24 14:17:17 --	2005-09-25 17:27:19&	42.8 		\\
100037010&	359.75, 	$-$0.05	&2005-09-29 04:35:41 --	2005-09-30 04:29:19&	43.7		\\
503099010&	359.78, 	1.13 		&2009-03-10 19:39:08 --	2009-03-11 10:56:14&	29.7 		\\
502008010&	359.83, 	0.67 		&2007-10-12 09:52:59 --	2007-10-12 21:50:19&	23.8 		\\
501046010&	359.83, 	0.33 		&2007-03-10 15:03:10 --	2007-03-11 03:55:14&	25.2 		\\
502003010&	359.83, 	$-$0.67 	&2007-10-10 03:41:13 --	2007-10-10 15:20:24&	21.5 		\\
502005010&	359.83, 	$-$1.00 	&2007-10-11 01:01:17 --	2007-10-11 11:32:20&	20.6 		\\
501008010&	359.85, 	$-$0.19 	&2006-09-26 14:18:16 --	2006-09-29 21:25:14&	129.6 	\\
501009010&	359.93, 	0.18 		&2006-09-29 21:26:07 --	2006-10-01 06:55:19&	51.2 		\\
504089010&	359.95, 	$-$1.20 	&2009-10-09 04:05:59 --	2009-10-10 14:10:06&	55.3 		\\
505079010&	359.95, 	$-$2.80 	&2011-03-12 06:36:20 --	2011-03-13 10:17:21&	50.2 		\\
505080010&	359.95, 	$-$3.80 	&2010-04-07 17:15:10 --	2010-04-09 21:14:16&	56.1 		\\
503103010&	359.99, 	1.20 		&2009-03-11 10:56:59 --	2009-03-11 19:17:08&	18.3 		\\
503008010&	0.00, 	$-$0.38 	&2008-09-03 22:53:29 --	2008-09-05 06:56:19&	53.7 		\\
504088010&	0.00, 	$-$0.83 	&2009-10-14 11:30:56 --	2009-10-15 15:29:19&	47.2 		\\
502059010&	0.00, 	$-$2.00 	&2007-09-29 01:40:51 --	2007-10-02 14:10:16&	136.8 	\\
503081010&	0.04, 	$-$1.66 	&2009-03-09 15:41:50 --	2009-03-10 19:36:19&	59.2 		\\
100027010&	0.06, 	$-$0.07 	&2005-09-23 07:18:25 --	2005-09-24 11:05:19&	44.8 		\\
100037040&	0.06, 	$-$0.07	&2005-09-30 07:43:01 --	2005-10-01 06:21:24&	43		\\
504050010&	0.10, 	$-$1.42 	&2010-03-06 03:55:37 --	2010-03-08 21:26:19&	100.4 	\\
502007010&	0.17, 	0.67 		&2007-10-11 23:09:15 --	2007-10-12 09:52:14&	22.0 		\\
502006010&	0.17, 	0.33 		&2007-10-11 11:34:01 --	2007-10-11 23:07:14&	22.6 		\\
502002010&	0.17, 	$-$0.67 	&2007-10-09 16:40:54 --	2007-10-10 03:40:24&	23.2 		\\
502004010&	0.17, 	$-$1.00 	&2007-10-10 15:21:17 --	2007-10-11 01:00:24&	19.9 		\\
502022010&	0.23, 	$-$0.27 	&2007-08-31 12:33:33 --	2007-09-03 19:00:25&	134.8 	\\
503007010&	0.33, 	0.17 		&2008-09-02 10:15:27 --	2008-09-03 22:52:24&	52.2 		\\
500005010&	0.43, 	$-$0.12 	&2006-03-27 23:00:22 --	2006-03-29 18:12:15&	88.4 		\\
100037060&	0.64, 	$-$0.10 	&2005-10-10 12:28:01 --	2005-10-12 07:05:23&	76.6 		\\
100037070&	1.00, 	$-$0.10 	&2005-10-12 07:10:24 --	2005-10-12 11:05:24&	9.2 		\\
501059010&	1.17, 	0.00 		&2007-03-15 18:55:51 --	2007-03-17 05:06:19&	62.2 		\\
501058010&	1.30, 	0.20 		&2007-03-14 05:02:29 --	2007-03-15 18:55:14&	63.3 		\\
501060010&	1.50, 	0.00 		&2007-03-17 05:07:04 --	2007-03-18 20:58:14&	64.8 		\\
508075010& 	1.75,		$-$0.04 	&2014-03-10 01:33:32 --  2014-03-12 15:30:12& 	109.3 	\\
502009010&	1.83, 	0.00 		&2007-10-12 21:52:24 --	2007-10-13 07:30:19&	20.9 		\\
505053010&	1.87, 	0.32 		&2011-03-23 03:51:35 --	2011-03-25 06:51:20&	100.9 	\\
507069010& 	2.00,		$-$0.04 	&2013-03-15 09:48:19 --  2013-03-17 18:39:15& 	110.3 	\\
507070010& 	2.25,		$-$0.04 	&2013-03-17 18:39:56 --  2013-03-20 02:40:03& 	111.8 	\\
507071010& 	2.50,		$-$0.04 	&2013-03-20 02:41:04 --  2013-03-22 07:19:07& 	112.3 	\\
507072010& 	2.75,		$-$0.04 	&2013-03-22 07:20:36 --  2013-03-24 08:45:10& 	110.7 	\\
507073010& 	3.00,		$-$0.04 	&2013-03-24 08:46:03 --  2013-03-26 09:53:12& 	108.9 	\\
508076010& 	3.25,		$-$0.04 	&2014-02-28 12:46:16 --  2014-03-02 17:00:14& 	109.8 	\\
508077010& 	3.50,		$-$0.04 	&2014-03-02 17:00:51 --  2014-03-04 23:00:15& 	109.4 	\\
503027010& 	5.72,		$-$0.06 	&2008-04-07 00:21:13 --  2008-04-07 16:30:23& 	31.0 		\\
503026010& 	5.89,		$-$0.38 	&2008-04-06 07:34:41 --  2008-04-07 00:20:24& 	31.7 		\\
500008010&	8.04, 	$-$0.05 	&2006-04-07 11:49:16 --	2006-04-08 10:54:18&	40.7 		\\
407092010&	8.14,		0.19 		&2012-09-21 08:05:31 --  2012-09-22 04:09:13& 	32.0 		\\
500007010&	8.44, 	$-$0.05 	&2006-04-06 14:41:18 --	2006-04-07 11:48:23&	37.5 		\\
401092010& 	9.94,		$-$0.27 	&2006-09-09 22:13:43 --  2006-09-11 04:00:14& 	48.9 		\\
402094010& 	9.95,		$-$0.27 	&2007-10-14 05:35:49 --  2007-10-15 08:00:23& 	52.2 		\\
406069010&	10.00,	$-$0.23	&2012-03-24 10:47:46 --  2012-03-26 12:45:20&	70.6		\\
504079010&	10.72, 	0.33 		&2009-09-11 17:59:44 --	2009-09-13 03:45:16&	51.0 		\\
503079010& 	10.84,	0.04 		&2008-04-01 16:33:52 --  2008-04-02 14:47:18& 	44.2 		\\
503078010& 	11.03,	0.07 		&2008-03-31 14:05:55 --  2008-04-01 16:30:23& 	51.5 		\\
504078010&	11.33, 	$-$0.06 	&2009-09-10 11:36:47 --	2009-09-11 17:58:19&	52.5 		\\
504077010&	11.61, 	$-$0.25 	&2009-09-09 04:55:29 --	2009-09-10 11:35:14&	51.9 		\\
502001010&	11.95, 	$-$0.09 	&2007-10-02 14:17:21 --	2007-10-03 23:40:19&	53.8 		\\
503087010& 	12.82,	$-$0.02 	&2009-03-04 19:45:39 --  2009-03-06 02:31:19& 	56.2 		\\
401101010& 	12.87,	0.01 		&2007-03-01 21:35:58 --  2007-03-03 06:45:19& 	63.8 		\\
502053010&	15.82, 	$-$0.84 	&2007-10-07 02:16:29 --	2007-10-08 18:10:14&	71.5 		\\
503030010& 	17.47,	$-$0.58 	&2008-10-19 04:41:55 --  2008-10-20 19:00:07& 	55.5 		\\
503028010& 	17.61,	$-$0.84 	&2008-10-15 21:49:50 --  2008-10-17 11:00:23& 	57.2 		\\
503029010& 	17.73,	$-$0.44 	&2008-10-17 11:01:12 --  2008-10-19 04:41:14& 	57.2 		\\
501044010& 	17.87,	$-$0.70 	&2006-10-17 19:37:16 --  2006-10-19 04:02:15& 	50.3 		\\
503086010& 	18.00,	$-$0.69 	&2009-03-19 21:33:25 --  2009-03-21 01:56:19& 	52.1 		\\
501045010& 	18.44,	$-$0.84 	&2006-10-19 04:03:16 --  2006-10-20 12:10:25& 	52.2 		\\
506051010& 	18.78,	0.40 		&2012-03-08 22:01:58 --  2012-03-10 03:50:15& 	52.0 		\\
507044010& 	19.57,	0.01 		&2012-10-15 13:05:48 --  2012-10-19 17:57:06& 	171.8 	\\
505025010& 	22.00,	0.00 		&2010-04-16 14:27:26 --  2010-04-17 17:27:12& 	50.5 		\\
506021010&	23.29, 	0.30 		&2011-04-08 06:06:16 --	2011-04-09 08:17:25&	40.3 		\\
904006010& 	23.40,	0.04 		&2010-03-27 09:03:32 --  2010-03-28 11:37:12& 	42.3 		\\
505026010& 	23.49,	0.04 		&2010-10-20 13:34:39 --  2010-10-22 01:45:11& 	49.0 		\\
401026010&	25.21, 	$-$0.12 	&2007-03-05 12:49:14 --	2007-03-06 10:17:14&	42.2 		\\
504099010&	25.50, 	0.00 		&2009-04-06 02:57:46 --	2009-04-07 17:52:14&	52.7 		\\
505088010&	26.30, 	0.00 		&2011-03-25 07:00:01 --	2011-03-26 10:40:15&	49.7 		\\
505089010& 	26.40,	$-$0.31 	&2011-03-26 10:41:12 --  2011-03-27 15:20:23& 	50.0 		\\
504052010&	26.44, 	0.13 		&2009-04-13 15:32:05 --	2009-04-14 18:11:19&	41.1 		\\
505090010& 	26.71,	$-$0.15 	&2011-03-27 15:21:16 --  2011-03-28 19:00:18& 	49.6 		\\
505091010& 	27.13,	$-$0.28 	&2011-03-28 19:01:23 --  2011-03-29 23:07:13& 	51.3 		\\
500009010&	28.46, 	$-$0.20 	&2005-10-28 02:40:08 --	2005-10-30 21:30:15&	93.3 		\\
500009020&	28.46, 	$-$0.20	&2006-10-15 02:15:12 --	2006-10-17 19:32:19&	98.9		\\
404081010&	29.71, 	$-$0.24 	&2009-04-15 19:37:17 --	2009-04-18 16:16:14&	104.3 	\\
508022010& 	35.61,	$-$0.40 	&2013-10-28 23:25:55 --  2013-10-30 03:09:09& 	52.6 		\\
506019010& 	36.00,	0.05 	 	&2011-09-18 22:04:29 --  2011-09-19 23:37:16& 	40.9 		\\
505027010&	37.00, 	$-$0.10 	&2010-04-17 17:35:13 --	2010-04-18 21:09:18&	51.0 		\\
509038010& 	39.19,	$-$0.30 	&2014-04-26 06:42:08 --  2014-04-28 02:56:13& 	82.8 		\\
\hline
\end{longtable}

Suzaku observations of the GDXE were carried out with the X-ray Imaging Spectrometers (XIS, \cite{Koyama2007a}) 
placed at the focal planes of 
the thin-foil X-ray Telescopes (XRT, \cite{Serlemitsos2007}).
The XIS consisted  of 4 sensors: XIS sensor-1 (XIS1) had a back-illuminated CCD (BI), 
while the other three XIS sensors (XIS0, 2, and 3) had  front-illuminated CCDs (FI).
Since XIS\,2 turned dysfunctional on 2006 November 9,  the other three sensors (XIS\,0, 1, and 3) were operated after the epoch.
A small fraction of the XIS 0 area was not  used since 2009 June 23 because of the  damage by a possible micro-meteorite.
The XIS was operated in the normal clocking mode. The field of view (FOV) of the XIS was \timeform{17'.8}$\times$\timeform{17'.8}.

We selected  the data set near the Galactic plane from all the Archive Suzaku data, where no bright X-ray source was included. 
The number of the data set (pointing positions) was 143, about 2.3 times larger than that of the previous work by \citet{Uchiyama2013}.
The pointing positions (Galactic coordinates) and exposure times are  listed in table 1. 

Data reduction and analysis were made using the HEASOFT.
The XIS pulse-height data for each X-ray event were converted to Pulse Invariant (PI) channels using the {\tt xispi} software 
and the calibration database.
We excluded the data obtained at the South Atlantic Anomaly, during Earth occultation, and at low elevation angles 
from the Earth rim of $<$ 5$^{\circ}$ (night Earth) or $<20^{\circ}$ (day Earth). 
After removing hot and flickering pixels, we used the grade 0, 2, 3, 4, and 6 data. 

\section{Analysis and Resluts} 

\subsection{Derivations  of the  Fe\,\emissiontype{I}\,K$\alpha$,
Fe\,\emissiontype{XXV}\,He$\alpha$ and
Fe\,\emissiontype{XXVI}\,Ly$\alpha$ fluxes} 

\begin{figure*}
        \begin{center}
        \includegraphics[width=8cm]{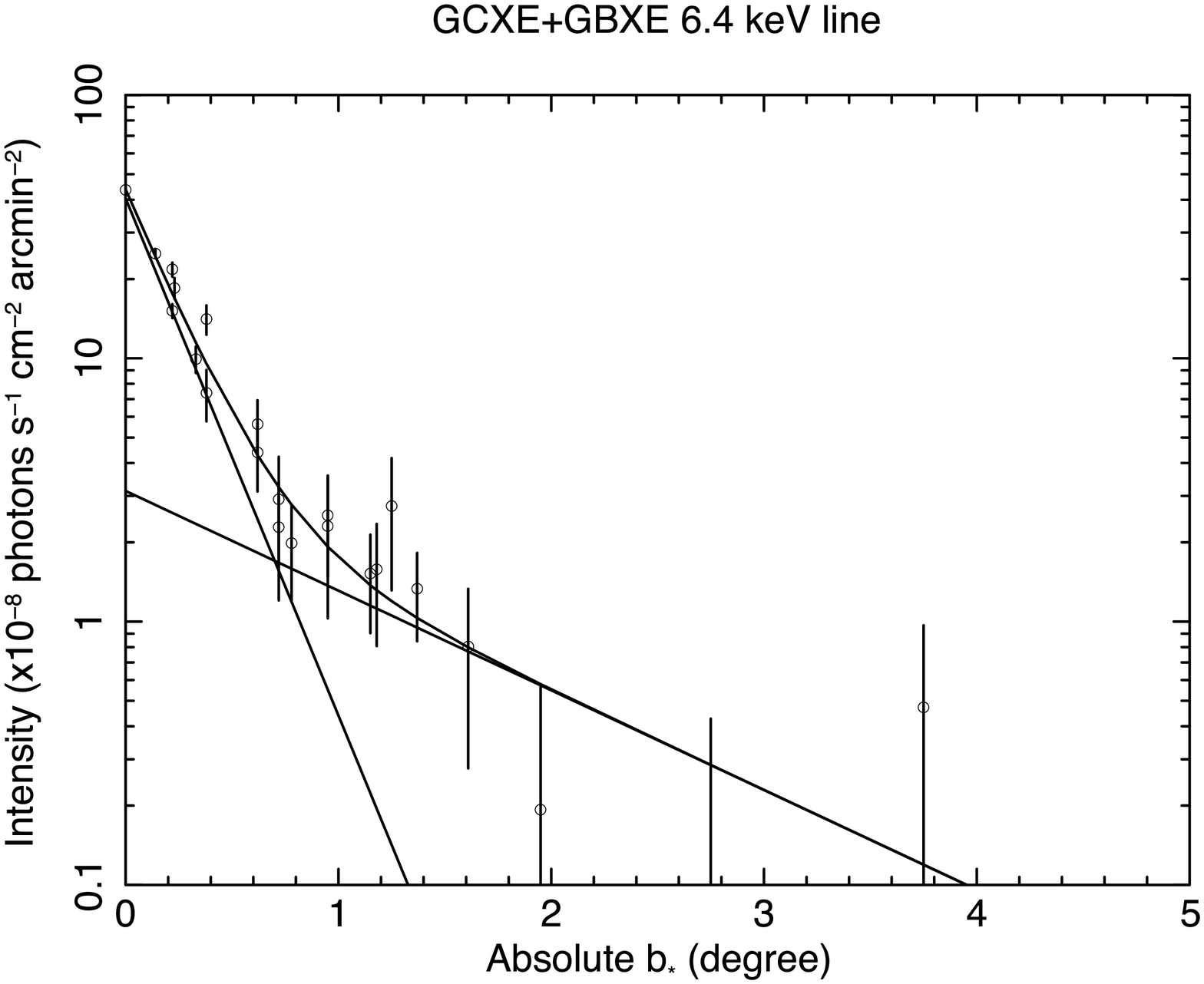}
        \includegraphics[width=8cm]{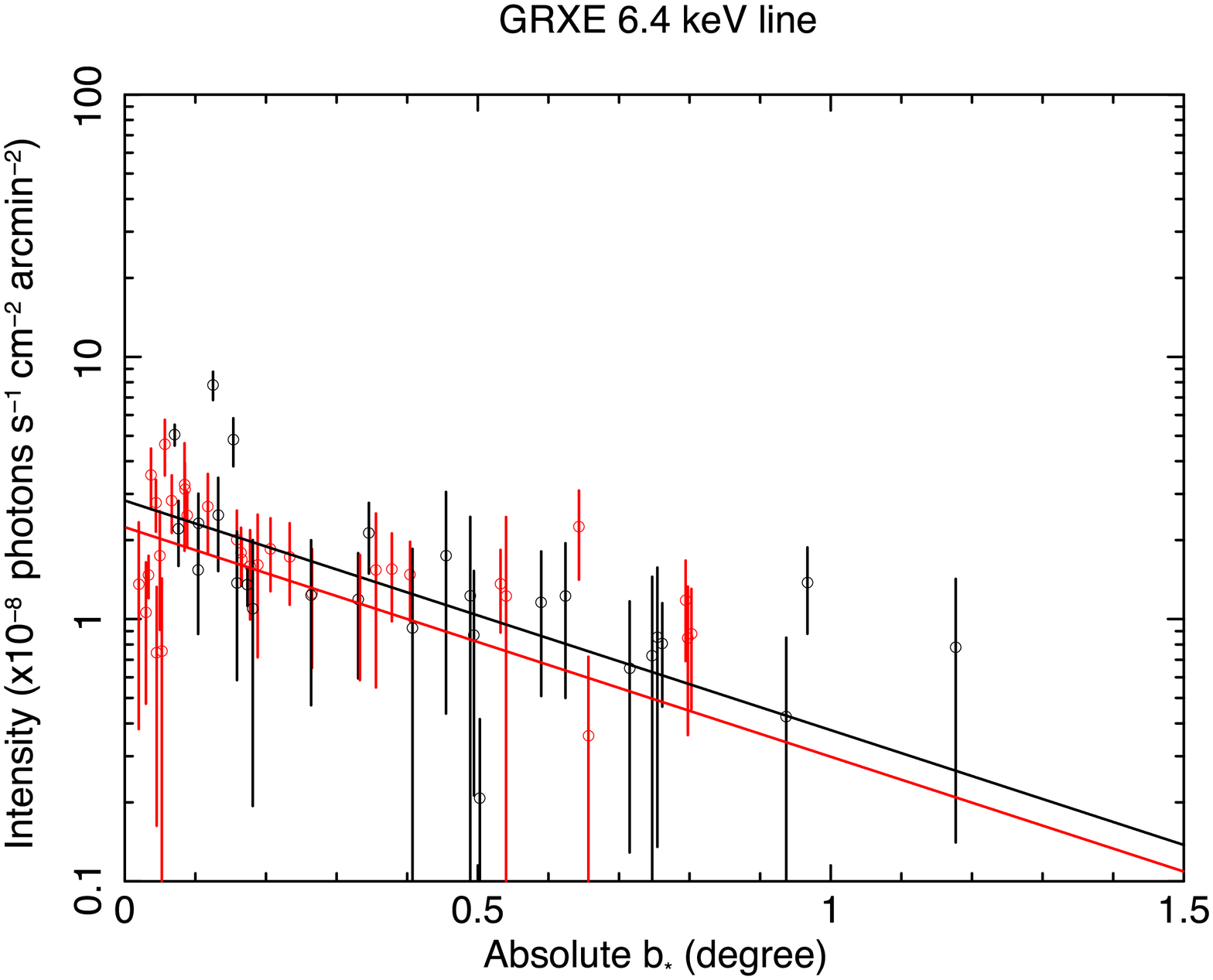}
        \includegraphics[width=8cm]{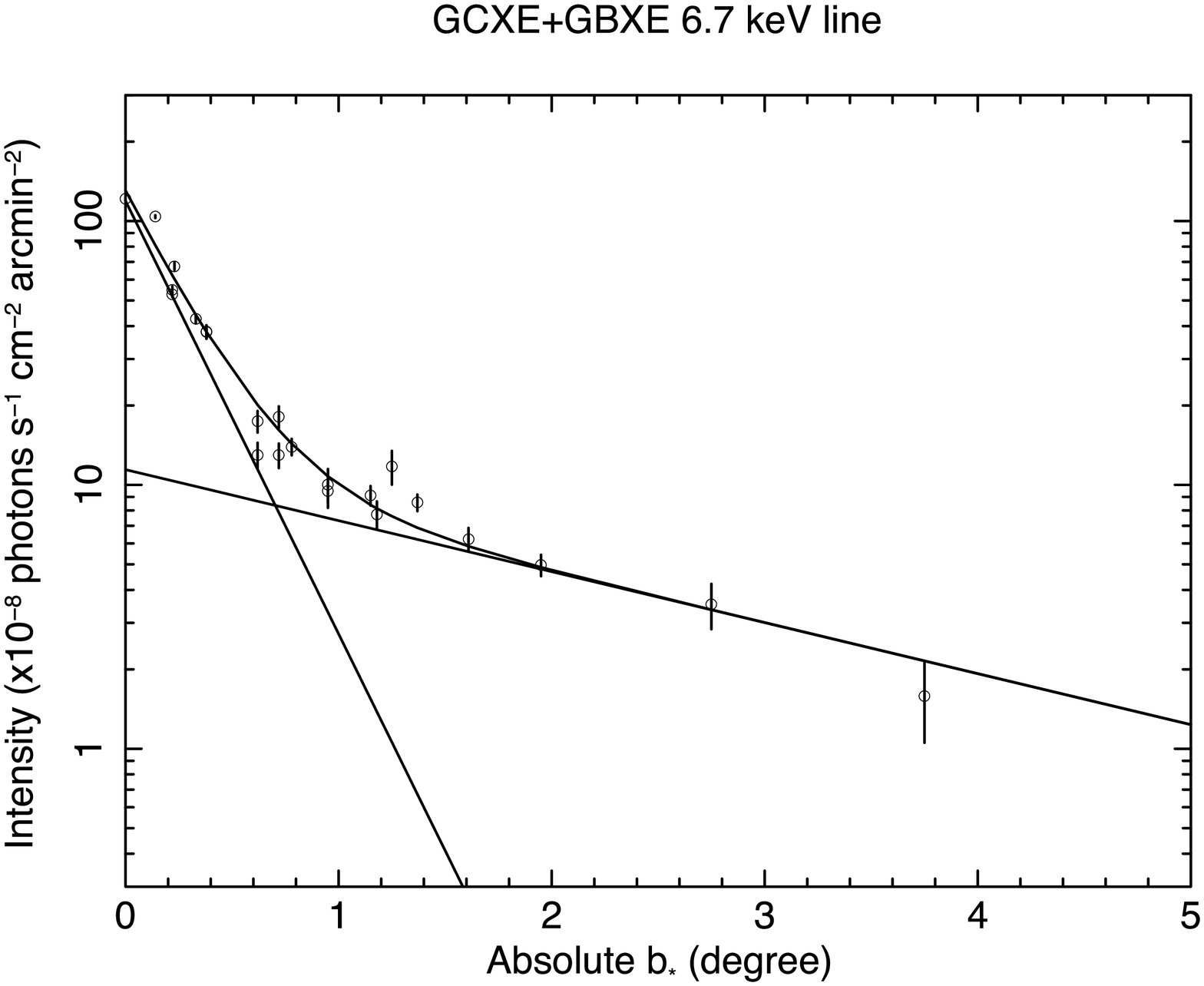}
        \includegraphics[width=8cm]{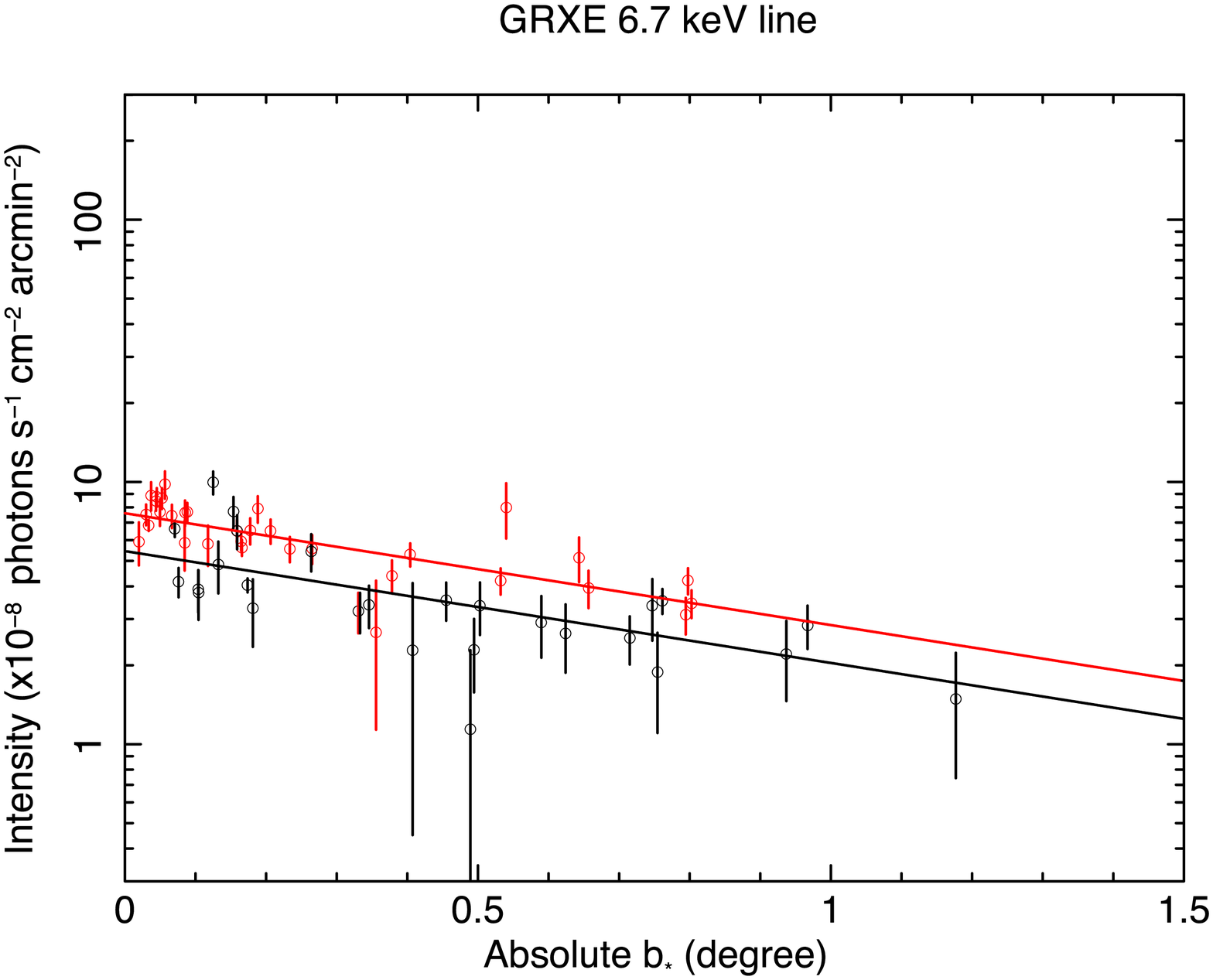}
        \includegraphics[width=8cm]{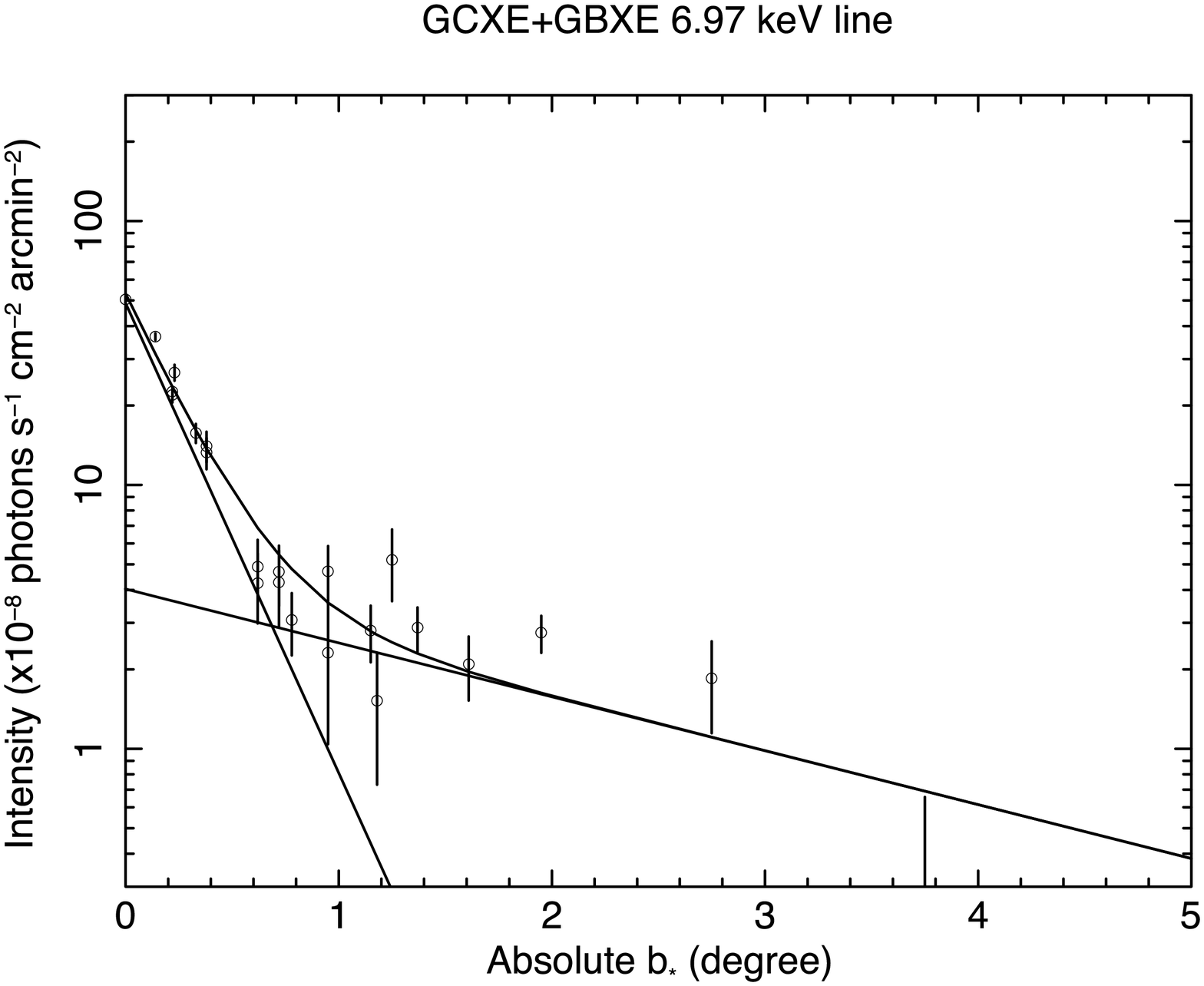}
        \includegraphics[width=8cm]{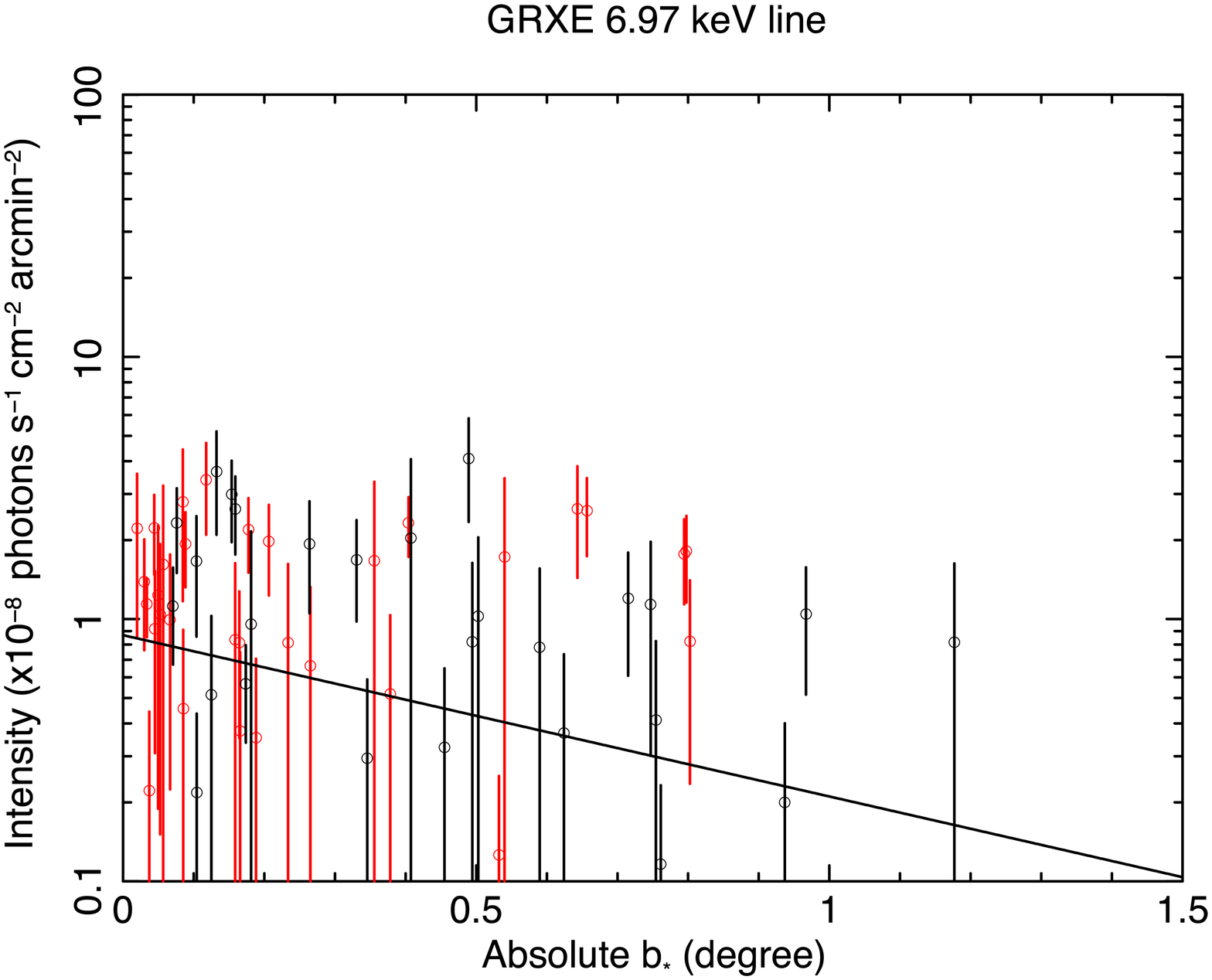}
        \end{center}
\caption{Galactic latitude distribution  of the Fe\,\emissiontype{I}\,K$\alpha$ (6.4 keV) (top), 
Fe\,\emissiontype{XXV}\,He$\alpha$ (6.7 keV) (middle) and  Fe\,\emissiontype{XXVI}\,Ly$\alpha$ (6.97 keV) (bottom) line fluxes.
Left: region (a) data of $|l|<$\timeform{0.D5}. The black lines show the best-fit model for the GCXE and GBXE. 
Right: region (d) data from $l$=\timeform{10D}--\timeform{30D} (red) and $l$=\timeform{330D}--\timeform{350D} (black). 
The black and the red lines show the best-fit model.}
\end{figure*}

We extracted X-ray photons from the entire region of the XIS FOV,
excluding discrete sources in the FOV and the calibration sources
located at the corners of the XIS sensors. 
In order to achieve the highest signal-to-noise ratio in the Fe band, we used only the FI detectors 
because the sensitivity in the Fe band was better than that of the BI detector \citep{Koyama2007a}. 
To maximize the photon statistics, data of each XIS sensor were merged.
The response files, Redistribution Matrix Files (RMFs) and Ancillary Response Files (ARFs), were made for each data set
using {\tt xisrmfgen} and {\tt xissimarfgen} of the HEASOFT package, respectively.
The non-X-ray background (NXB) was constructed from the night earth data provided by the XIS team using  {\tt xisnxbgen} of 
the HEASOFT package \citep{Tawa2008}.  

We made an X-ray spectrum in the 4--10 keV band from each position. 
After the subtraction of the NXB, we fitted the spectra with a phenomenological model: 
a power-law plus a bremsstrahlung and many Gaussian lines. 
The power-law is the cosmic X-ray background (CXB) with the fixed photon index ($\Gamma$) and flux of 1.4 and 
10 photons s$^{-1}$ cm$^{-2}$ sr$^{-1}$ keV$^{-1}$ at 1 keV, respectively 
\citep{Marshall1980, Gendreau1995,Kushino2002,Revnivtsev2005}.
The bremsstrahlung and Gaussian lines are for the GDXE model.

We assumed the absorption column for the GC regions of $|l|\le$\timeform{5D} and $|b|\le$\timeform{0.D5}, 
$N_{\rm H}$(GCXE), to be 6$\times10^{22}$~cm$^{-2}$ \citep{Sakano2002}.
For the GRXE and GBXE regions of $|b|\le$\timeform{1D} and $|b|\ge$\timeform{1D}, 
$N_{\rm H}$(GRXE) and $N_{\rm H}$(GBXE) were fixed to 3$\times10^{22}$~cm$^{-2}$ and 1$\times10^{22}$~cm$^{-2}$, respectively. 
We note that the assumed $N_{\rm H}$ has no large effect in the energy band of 5--8 keV. 
The absorption of the CXB, $N_{\rm H}$(CXB), was assumed to be twice of the  interstellar absorption of $N_{\rm H}$(GDXE). 
The cross section of photoelectric absorption was taken from Balucinska-Church and McCammon (1992).
As noted in \citet{Koyama2007b}, in the GDXE spectrum, a clear absorption edge of neutral or lower ionized iron was found at 7.1 keV. 
Therefore we set the Fe abundance of the absorption column 
as a free parameter if the spectra exhibited a deep absorption edge.

The temperature and the normalization of the bremsstrahlung were free parameters. 
The fluxes of the Fe\,\emissiontype{I}\,K${\alpha}$ (6.4 keV), Fe\,\emissiontype{XXV}\,He${\alpha}$ (6.7 keV) 
and Fe\,\emissiontype{XXVI}\,Ly${\alpha}$ (6.97 keV) lines were also free parameters, but the flux of the Fe\,\emissiontype{I} K${\beta}$ line 
at 7.058 keV was  fixed to the theoretical value of 0.125 times  
Fe\,\emissiontype{I}\,K${\alpha}$ \citep{Kaastra1993}.
Since the Fe\,\emissiontype{XXV}\,He${\alpha}$ line was a blend of the resonance, inter-combination and forbidden lines, 
the intrinsic line width of Fe\,\emissiontype{XXV}\,He${\alpha}$ was assumed to be 23~eV \citep{Koyama2007b}.
Emission lines of Ni~\emissiontype{I}\,K$\alpha$ (7.49 keV),
Ni~\emissiontype{XXVII}\,He$\alpha$ (7.77 keV), 
Fe\,\emissiontype{XXV}\,He$\beta$ (7.88 keV), 
Fe\,\emissiontype{XXVI}\,Ly$\beta$ (8.25 keV), 
Fe\,\emissiontype{XXV}\,He$\gamma$ (8.29 keV) 
and Fe\,\emissiontype{XXVI} Ly$\gamma$ (8.70 keV) were added 
if the spectra had high statistics.

\subsection{Scale height} 

\begin{table*}[htb] 
\caption{Best-fit parameters of the GCXE, GBXE and GRXE.}
\begin{center}\begin{tabular}{llccccc} \hline 
Region & Component & \multicolumn{5}{c}{Parameter} \\
&&\multicolumn{3}{c}{Normalization ($A^\ast$)} &e-folding scale ($b^\dag$) &Scale height$^\ddag$ \\
\hline
&&$l$=\timeform{0D}&$l$=\timeform{358.D5}& $l$=\timeform{356.D0}--\timeform{356.D4}& &\\
\hline
GCXE&6.4~keV&4.1$\pm$0.2& = $A_{l=0^{\circ}}$$\times$0.11 
& = $A_{l=0^{\circ}}$$\times$0.004  & 0.22$\pm$0.02&31$\pm$3\\
     &6.7~keV&11.9$\pm$0.6& = $A_{l=0^{\circ}}$$\times$0.11& = $A_{l=0^{\circ}}$$\times$0.004 & 0.26$\pm$0.02&36$\pm$3 \\
     &6.97~keV&4.9$\pm$0.2& = $A_{l=0^{\circ}}$$\times$0.11& = $A_{l=0^{\circ}}$$\times$0.004 &0.24$\pm$0.02&34$\pm$3 \\
     &5--8~keV&77$\pm$4& = $A_{l=0^{\circ}}$$\times$0.11& = $A_{l=0^{\circ}}$$\times$0.004 & 0.25$\pm$0.02&35$\pm$3 \\
             \hline
GBXE  & 6.4 keV & 0.31$\pm$0.15 & 0.35$\pm$0.10 & 0.28$\pm$0.07 & 1.15$\pm$0.36&161$\pm$50\\
             & 6.7 keV & 1.14$\pm$0.34 & 1.15$\pm$0.27 & 1.04$\pm$0.21 & 2.25$\pm$0.68&314$\pm$95\\
             & 6.97 keV & 0.40$\pm$0.12 & 0.39$\pm$0.10 & 0.19$\pm$0.06 & 2.13$\pm$0.66&297$\pm$92\\
             & 5--8 keV & 12$\pm$2& 10.6$\pm$1.4& 7.2$\pm$0.9 & 1.96$\pm$0.25&274$\pm$35\\
             \hline
&&$l$=\timeform{10D}--\timeform{30D}&$l$=\timeform{330D}--\timeform{350D}& & & \\
\hline
GRXE  & 6.4 keV & 0.23$\pm$0.03& 0.28$\pm$0.04& & 0.50$\pm$0.12&70$\pm$17\\
       & 6.7 keV & 0.76$\pm$0.02& 0.54$\pm$0.03& & 1.02$\pm$0.12&142$\pm$17\\
       & 6.97 keV & 0.09$\pm$0.02& = $A_{l=\timeform{10D}-\timeform{30D}}$ & & 0.71$\pm$0.29&99$\pm$40\\
         & 5--8 keV & 5.8$\pm$0.4 & 4.9$\pm$0.5& & 1.04$\pm$0.20&145$\pm$28\\
\hline 
\end{tabular} 
\end{center}
Error is 1 $\sigma$ (68\% confidence) level. \\
$\ast$:~Unit is 10$^{-7}$ photons s$^{-1}$ cm$^{-2}$ arcmin$^{-2}$.\\
$\dag$:~Unit is degree.\\
$\ddag$:~Unit is pc. Distance of 8 kpc is assumed.\\
\end{table*}

In order to investigate the latitude distribution of the Fe\,\emissiontype{I}\,K$\alpha$, Fe\,\emissiontype{XXV}\,He$\alpha$ and 
Fe\,\emissiontype{XXVI}\,Ly$\alpha$ lines, and the 5--8 keV band flux,
the best-fit results were  grouped into the 4 regions: 
(a) $|l|<$\timeform{0.D5}, (b) $l$=\timeform{358.D5},
(c) $l$=\timeform{356.D0}--\timeform{356.D4} and 
(d) $|l|$=\timeform{10D}--\timeform{30D}.
Here and after, we used a new Galactic coordinate of ($l_{\ast}$, $b_{\ast}$)= ($l + \timeform{0.D056}$, $b + \timeform{0.D046}$), 
referring the GC (Sgr A$^{\ast}$) position of ($l$, $b$)=(\timeform{-0.D056}, \timeform{-0.D046}).

The latitude profiles of the Fe\,\emissiontype{I}\,K$\alpha$, Fe\,\emissiontype{XXV}\,He$\alpha$ 
and Fe\,\emissiontype{XXVI}\,Ly$\alpha$ lines in the regions (a) and (d) are given in figure 1. 
The left panels clearly show an existence of two components, while the right panels show a single component.
The region (a) is mainly the GCXE data with a  small fraction of the GBXE, while the regions (b) and (c) are vice versa. 
The region (d) is the data of the pure GRXE \citep{Yamauchi1993, Uchiyama2013}. 

We  simultaneously fitted the profiles of (a), (b) and (c) with a two-exponential model of
\begin{equation}
I(b_{\ast})=A_{\rm GCXE}\ {\rm exp}(-\frac{|b_{\ast}|}{b_{\rm GCXE}})+A_{\rm GBXE}\ {\rm exp}(-\frac{|b_{\ast}|}{b_{\rm GBXE}}), 
\end{equation}
where $b_{\rm GCXE}$ and $b_{\rm GBXE}$ are e-folding scales (degree) of the GCXE and GBXE, respectively 
and $A_{\rm GCXE}$ and $A_{\rm GBXE}$ are normalizations of the GCXE and GBXE, respectively.
We linked $b_{\rm GCXE}$ and $b_{\rm GBXE}$ of (a), (b) and (c) each other and 
scaled $A_{\rm GCXE}$ of (b) and (c) to (a) using the e-folding longitude scale of the GCXE of \timeform{0.D63} 
\citep{Uchiyama2013}.

The data of (d) were fitted with a one-exponential model of
\begin{equation}
I(b_{\ast})=A_{\rm GRXE}\ {\rm exp}(-\frac{|b_{\ast}|}{b_{\rm GRXE}}),
\end{equation}
where $b_{\rm GRXE}$ and $A_{\rm GRXE}$ are the e-folding scale (degree) 
and normalization of the GRXE, respectively.
We excluded the local enhanced regions, the XRNe and bright supernova remnant (SNR) Sgr A East in the GCXE 
(e.g., \cite{Koyama1996,Park2004}).

The best-fit parameters are listed in table 2. 
The e-folding scales in table 2 are essentially the same as those derived by \citet{Uchiyama2013}, 
except for the e-folding latitude scale of the GRXE. 
This disagreement was due to the data set selection; \citet{Uchiyama2013} used the data mainly near the GCXE, 
and hence the e-folding scale of the GRXE was largely affected by the large value of the GBXE (see table 2).  
Our estimate of the GRXE was based on  a lot of pure GRXE results in the range of $|l|=\timeform{10D}$--$\timeform{30D}$, 
and hence would be more reliable.  
On the other hand, the e-folding longitude scale for the GCXE and GRXE by \citet{Uchiyama2013} 
would be reliable due to limited contribution of the GBXE.   
Assuming the distance of 8 kpc, the e-folding scales (degree) of the latitude distribution in the 6-th column of table 2 were 
converted to the SHs (pc). The results are listed in the last column in table 2.  

\subsection{Equivalent width} 

Figure 2 show the longitude profiles of the line flux of Fe\,\emissiontype{I}\,K$\alpha$,
Fe\,\emissiontype{XXV}\,He$\alpha$ and Fe\,\emissiontype{XXVI}\,Ly$\alpha$, and those of the flux ratios of 
Fe\,\emissiontype{I}\,K$\alpha$/Fe\,\emissiontype{XXV}\,He$\alpha$ and
 Fe\,\emissiontype{XXVI}\,Ly$\alpha$/Fe\,\emissiontype{XXV}\,He$\alpha$.
The longitude distribution of the Fe\,\emissiontype{XXV}\,He$\alpha$
and Fe\,\emissiontype{XXVI}\,Ly$\alpha$ lines are symmetry with respect to the Galactic center. 
However the  Fe\,\emissiontype{I}\,K$\alpha$  flux  and  the flux ratio relative to the Fe\,\emissiontype{XXV}\,He$\alpha$ line 
(Fe\,\emissiontype{I}\,K$\alpha$/Fe\,\emissiontype{XXV}\,He$\alpha$) show east-west asymmetry at $l$=\timeform{1.D5}--\timeform{3.D5} and 
$l$= \timeform{330D}--\timeform{340D} regions (see figure 2, the 1-st and 4-th panels). 

\begin{figure}
  \begin{center}
        \includegraphics[width=8cm]{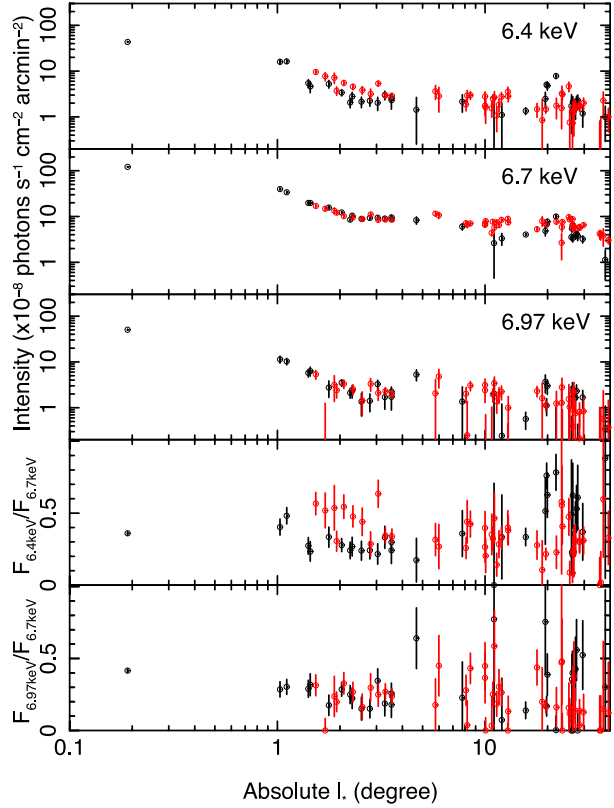}
  \end{center}
\caption{Galactic longitude distribution of the   Fe\,\emissiontype{I}\,K$\alpha$,
Fe\,\emissiontype{XXV}\,He$\alpha$ and Fe\,\emissiontype{XXVI}\,Ly$\alpha$ line fluxes,
and the flux ratios of  
Fe\,\emissiontype{I}\,K$\alpha$/Fe\,\emissiontype{XXV}\,He$\alpha$ and
 Fe\,\emissiontype{XXVI}\,Ly$\alpha$/Fe\,\emissiontype{XXV}\,He$\alpha$. 
Referring the e-folding scale of the GCXE and GRXE (table 2), we select the data of
$|b_{\ast}|<$\timeform{0.D2} and $|b_{\ast}|<$\timeform{0.D5} in the regions of $|l_{\ast}|<$\timeform{1.D5} (GCXE) 
and $|l_{\ast}|>$\timeform{1.D5} (GRXE), respectively. 
The data containing the XRNe and Sgr A East SNR are excluded.
The red and black colors show the data of $l_{\ast}>$\timeform{0D} and $l_{\ast}<$\timeform{0D}, respectively.
}
\label{fig:sample}
\end{figure}

We obtained the EWs of the Fe\,\emissiontype{I}\,K$\alpha$ (EW$_{6.4}$),
 Fe\,\emissiontype{XXV}\,He$\alpha$ (EW$_{6.7}$) and
Fe\,\emissiontype{XXVI}\,Ly$\alpha$ (EW$_{6.97}$) lines from the positions of 
the GCXE ($|l|<$\timeform{1.D5}, $|b|\le$\timeform{0.D5}), 
GBXE ($|l|<$\timeform{4.D0},  $|b|\ge$\timeform{1.D0}) 
and GRXE ($|l|$=\timeform{10D}--\timeform{30D}, $|b|\le$\timeform{1.D0}),  
where the local enhancements due to XRNe and the supernova remnant (SNR) Sgr A East in the GCXE were excluded. 
 
The EW$_{6.4}$, EW$_{6.7}$ and EW$_{6.97}$ relations of the GCXE are plotted in figure 3a and  3b. 
Although the EW$_{6.4}$ and EW$_{6.7}$ show no clear correlation (a correlation coefficient, R$\sim$0.1), 
the EW$_{6.97}$ and EW$_{6.7}$ show a correlation (R$\sim$0.6). 
The best-fit proportional line is plotted in figure 3b.
The same plots of the EW$_{6.4}$ and EW$_{6.7}$ relations of the GBXE and GRXE are shown in figure 3c and 3d, respectively. 
We also made  the EW$_{6.7}$ and EW$_{6.97}$ relation plots in the GBXE and GRXE. 
Due to the large statistical errors, we found no clear correlation in the GBXE and GRXE data (R$\sim$$-$0.2 -- 0.5).

For the GCXE, GBXE and GRXE, the mean EW$_{6.4}$ values were 145$\pm{3}$, 61$\pm{11}$ and 97$\pm{12}$~eV, the EW$_{6.7}$ were 
527$\pm{4}$, 443$\pm{14}$ and 428$\pm{15}$~eV
and the EW$_{6.97}$ were 221$\pm{3}$, 160$\pm{14}$ and 117$\pm{19}$~eV, respectively.

\begin{figure*} 
  \begin{center}
        \includegraphics[width=8cm]{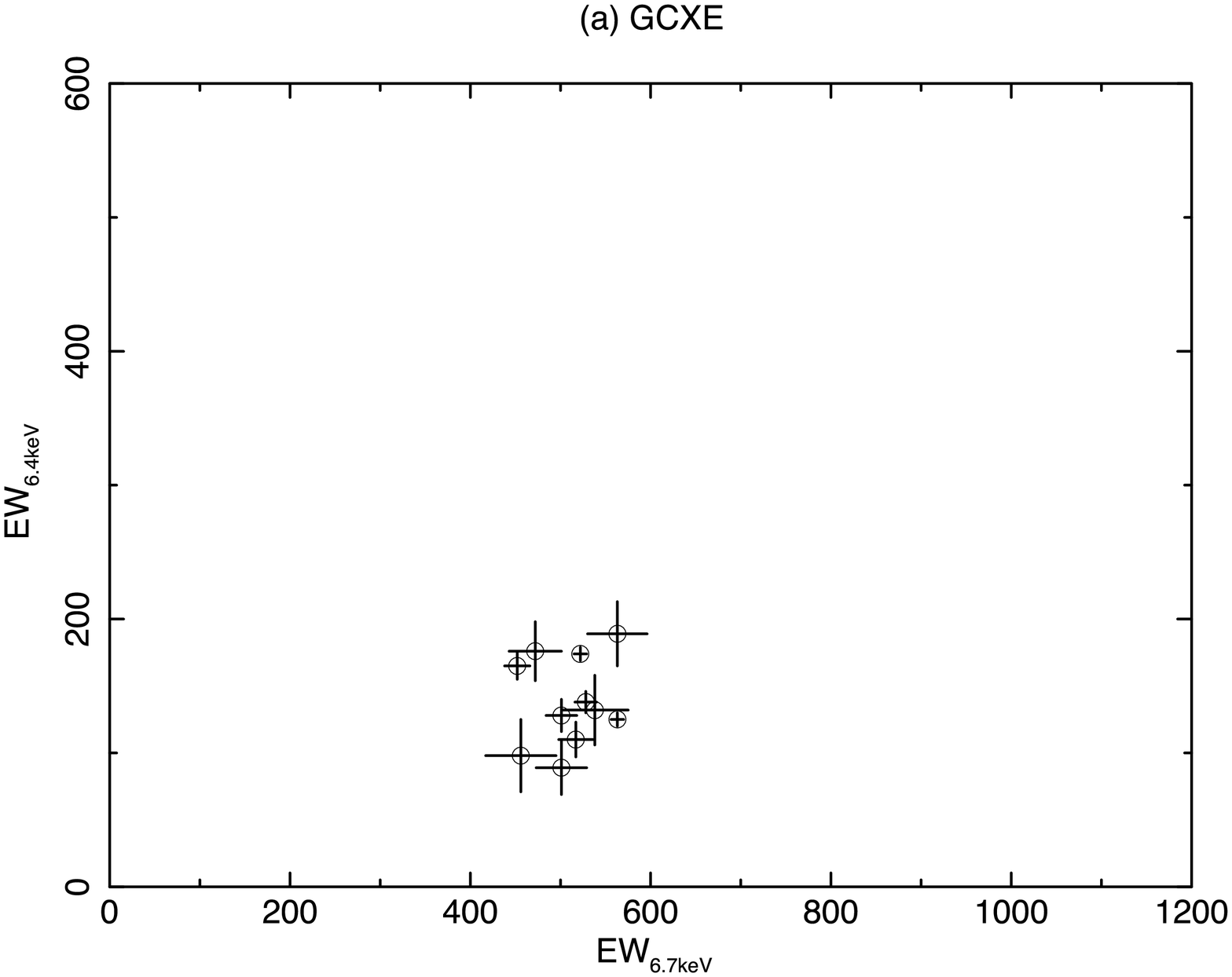}
        \includegraphics[width=8cm]{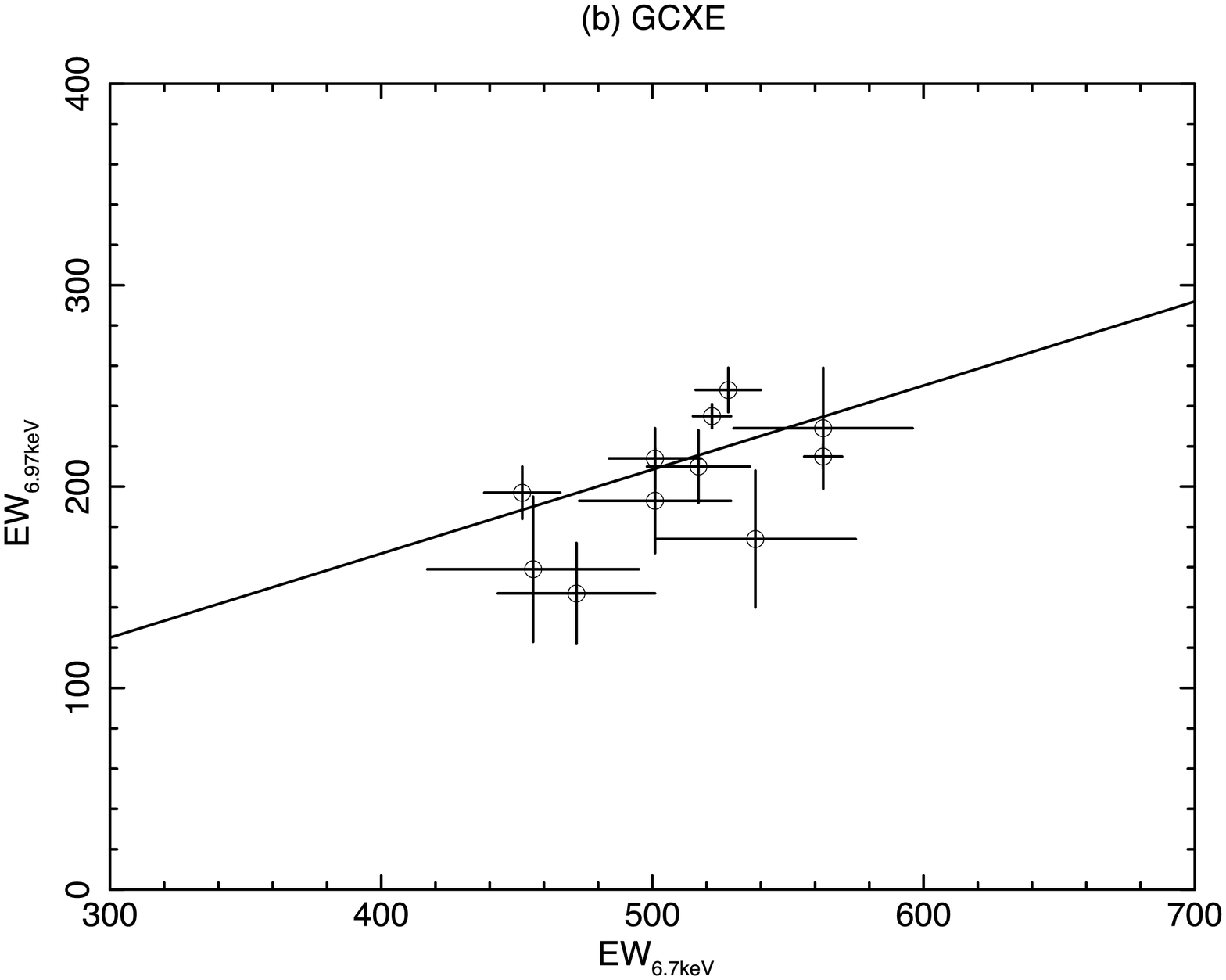}
        \includegraphics[width=8cm]{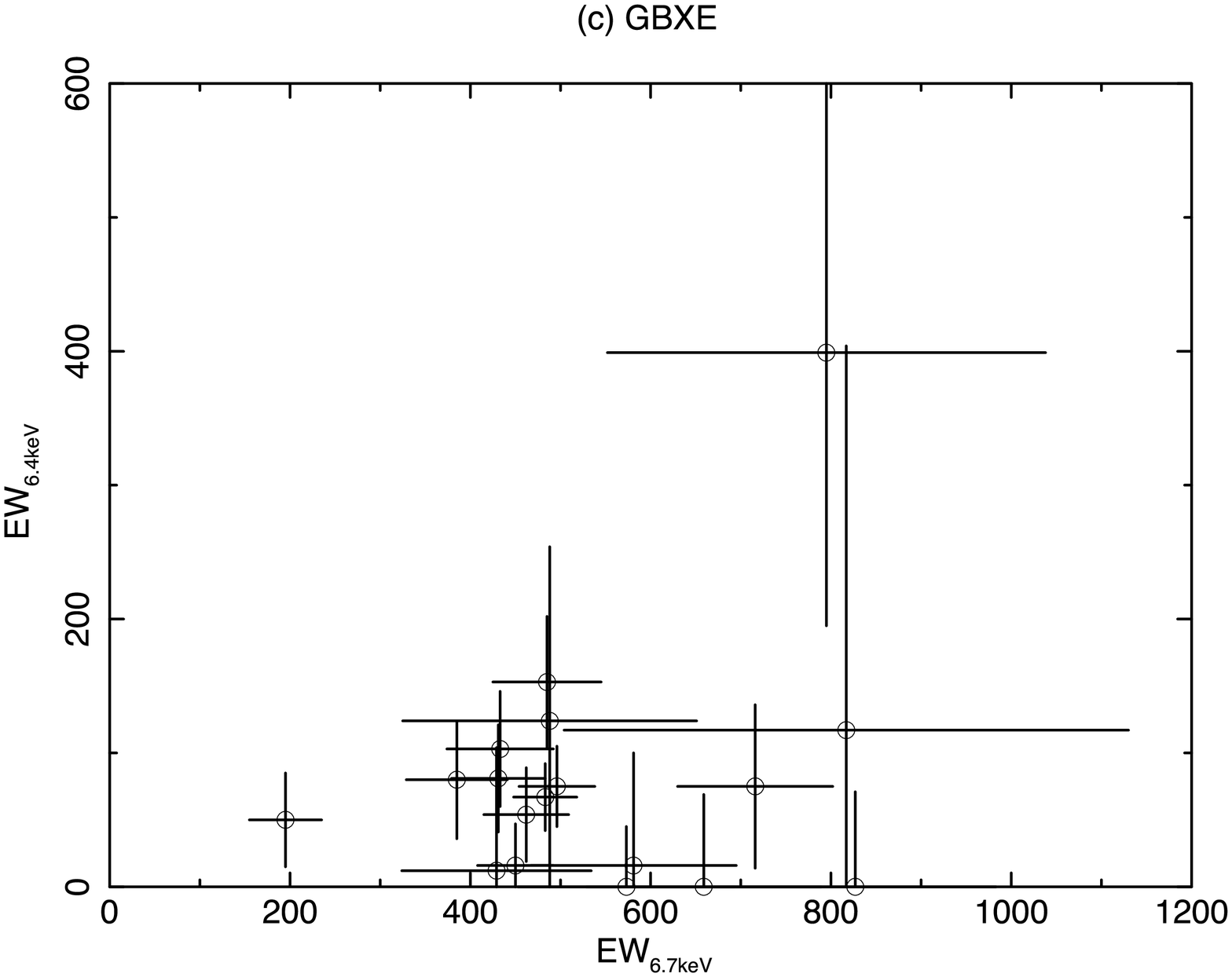}
        \includegraphics[width=8cm]{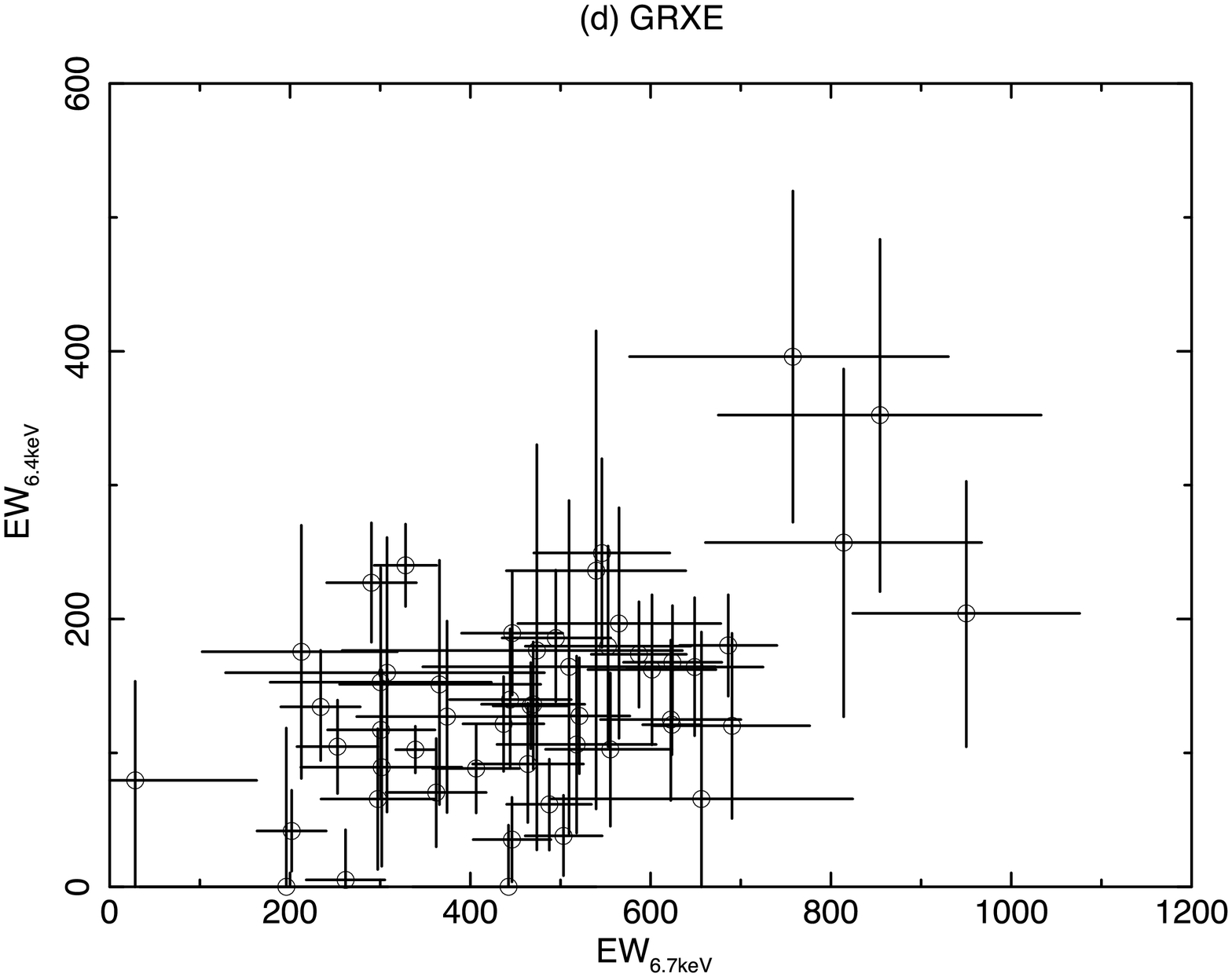}
  \end{center}
\caption{(a) Correlation plot of the EWs of EW$_{6.4}$ and EW$_{6.7}$ in the GCXE.
(b)the same as (a) but the correlation is  EW$_{6.97}$ between  EW$_{6.7}$.
The solid line is the best-fit proportional line of EW$_{6.97}$= 0.42$\times$EW$_{6.7}$.
(c) the same as (a) but in the GBXE. (d) the same as (a) but in the GRXE. 
}
\label{fig:sample}
\end{figure*}

\section{Discussion} 

\subsection {Overview of the point source origin of the GDXE} 

Since the discovery of strong iron K-shell lines in the GDXE, the origin of the GDXE becomes a long standing question. 
One of the most accepted idea is the point source origin \citep{Revnivtsev2006}. 
The point source origin is based on the X-ray luminosity function (XLF) of the continuum flux (e.g., the 2--10~keV band). 
In the luminosity range of $\sim10^{30}$--$10^{33}$ ~erg~s$^{-1}$, 
the XLF of point sources consists of mainly cataclysmic variables (CVs) and active binaries (ABs).  It is made using the RXTE sky survey and the ROSAT all sky survey, where the ROSAT flux in the  0.1--2.4 keV band were converted to the 2--10 keV band. 
This conversion process causes a large systematic error of $\ge$50\% \citep{Sazonov2006}. In the lower luminosity band 
$5\times10^{27}$--$\sim10^{30}$~erg~s$^{-1}$,  the XLF mainly consists of coronal active stars (CAs), where the systematic error would be even larger \citep{Sazonov2006}.

Nevertheless, the strongest support on the point source origin at this time  came from the deep observation with Chandra.  
\citet{Revnivtsev2009} resolved  $\sim80$\% of the GDXE flux into point sources  following the XLF of \citet{Sazonov2006}. 
Although the authors did not compare the iron K-shell line fluxes (EW) 
with those of CVs and ABs, they regarded that major sources of the Fe\,\emissiontype{I}\,K$\alpha$, Fe\,\emissiontype{XXV}\,He$\alpha$ and 
Fe\emissiontype{XXVI}\,Ly$\alpha$ lines are CVs and ABs following \citet{Sazonov2006} 
because these sources have been known as strong iron line emitters.

We have separately determined the EWs and SHs of the Fe\,\emissiontype{I}\,K$\alpha$, 
Fe\,\emissiontype{XXV}\,He$\alpha$ and Fe\,\emissiontype{XXVI}\,Ly$\alpha$ lines in the GCXE, GBXE and GRXE. 
In the next subsections, 
we therefore re-examine the point source origin  of the GCXE, GBXE and GRXE 
based on these  new observational results of the EWs and SHs, 
in comparison with those of the published results of CVs, ABs and CAs.  
In the current point source origin, magnetized CVs (mCVs) cover the energy range of $\gtsim10^{32}$~erg~s$^{-1}$. 
Non magnetized CVs (nmCVs), often called as dwarf novae, and bright ABs are in the range of $\sim10^{30}$--$10^{32}$~erg~s$^{-1}$. 
The lowest luminosity band of $\lesssim10^{31}$~erg~s$^{-1}$ is covered by faint ABs and CAs (e.g., \cite{Sazonov2006}).

\subsection {Iron line equivalent widths and scale heights of CVs, ABs and CAs} 

Since the EWs of Fe\,\emissiontype{XXV}\,He$\alpha$ and Fe\,\emissiontype{XXVI}\,Ly$\alpha$ show a correlation (figure 3b),  
and the SHs of these lines are similar (table 2), the origin would be the same. 
We therefore sum EW$_{6.7}$ and EW$_{6.97}$ (EW$_{6.7}$$+$EW$_{6.97}$) hereafter.
The mean EW$_{6.4}$ and EW$_{6.7}$$+$EW$_{6.97}$ of mCVs are $\sim$120~eV and $\sim$260~eV, respectively 
\citep{Ezuka1999,Hellier1998, Hellier2004, Bernardini2012, Eze2015,Xu2016}.
There is significant variation of the observed mean EWs from the author to author.  
We checked the author-to-author variations and found to be $\sim $40\% at most. 
The same order of uncertainty would be exist in the following estimation of the mean EWs in the other point sources.

Since nmCVs have lower flux but about 10 times larger space density than mCVs \citep{Patterson1984}, 
they would be important contributors to the GDXE in the energy range of $\sim10^{30}$--$10^{32}$~erg~s$^{-1}$.  
\citet{Mukai1993} reported that the sum of EW$_{6.4}$$+$EW$_{6.7}$$+$EW$_{6.97}$ was $\sim$780~eV, 
where unreasonably large EW samples were excluded.  
\citet{Byckling2010} reported that EW$_{6.4}$ was $\sim$90~eV, while \citet{Rana2006} reported that 
EW$_{6.4}$, EW$_{6.7}$ and EW$_{6.97}$ were  $\sim$60~eV, $\sim$260~eV and $\sim$85~eV, respectively. 
\citet{Xu2016} analyzed 16 samples in the Suzaku archive and found EW$_{6.4}\sim$62 eV, EW$_{6.7}\sim$438 eV 
and EW$_{6.97}\sim$95 eV.
Thus in average, EW$_{6.4}$ and EW$_{6.7}$$+$EW$_{6.97}$ of nmCVs are $\sim$70~eV and $\sim$530~eV, respectively.

\citet{Schmitt1990} compiled the Einstein data of X-ray stars. The major sources are
ABs and CAs in the luminosity range of $\sim10^{30}$--10$^{32}$~erg~s$^{-1}$ and $10^{27}$--$10^{30}$~erg~s$^{-1}$, respectively. 
Since the spectral information of ABs  in the iron K-shell band has been very limited so far, we estimate the EW 
from the observed Fe abundance and plasma temperature. 
The mean temperature and iron abundance are reported to be $\sim$3~keV, and $\sim$0.3 solar, respectively 
\citep{Tsuru1989, Doyle1991, Dempsey1993, Antunes1994,White1994,Gudel1999, Osten2000, Audard2003, Pandey2012}, and hence
the EW$_{6.4}$ is negligible, and the sum of EW$_{6.7}$ and EW$_{6.97}$ is estimated to be $\sim$650~eV.
Recently, \citet{Xu2016} obtained EW$_{6.4}\le$20 eV, EW$_{6.7}\sim$286 eV and EW$_{6.97}\sim$12 eV from the 4 Suzaku samples.
Thus, EW$_{6.4}$ and EW$_{6.7}$$+$EW$_{6.97}$ of ABs are $\le$20~eV and 300--650~eV, respectively.
The 6.4 keV line would be due to irradiation of the stellar photosphere by the coronal hard X-rays.

The EWs of CAs are even more unclear, but may be an important component in the luminosity range of $\lesssim10^{30}$~erg~s$^{-1}$ 
\citep{Sazonov2006}. 
Since X-rays from CAs are due to dynamo activity, 
young CAs in the pre-main sequence (PMS) and fast rotating CAs in an earlier phase are more active than old CAs in the main sequence (MS) \citep{Gudel2004}. 
\citet{Pandey2008} reported the temperature of a late type dwarf to be $\le$~1 keV. 
The temperature is too low to excite the iron K-shell lines, and hence old CAs would be ignored as the candidate of the GDXE origin. 
The young CAs (PMS) in the star forming regions of the $\rho$-Oph and the Orion nebula clouds have the X-ray luminosity 
and the mean temperature of $10^{28}$--$10^{31.5}$~erg~s$^{-1}$ and 2--3~keV, respectively 
\citep{Imanishi2003, Ozawa2005, Pr2008}. 
The Fe abundance is $\sim$0.2--0.4 solar. 
A fraction of young CAs in molecular clouds (MCs) show EW$_{6.4}\sim$100--400~eV  \citep{Takagi2002, Tsujimoto2005, Czesla2010}.
\citet{Tsujimoto2005} concluded that the 6.4 keV line arises from reflection of circumstellar disks.
However, these are very rare cases, and hence
the mean EWs of young CAs may be more or less similar to the ABs.
In order to help the re-examination of the point source origin, we list the EWs of the GDXE and candidate point sources in table 3. 

\begin{table*}[htb] 
\caption{Equivalent width of mCVs, nmCVs and ABs}
\begin{center}
\begin{tabular}{lcccl} 
\hline 
Sources&EW$_{6.4}$~(eV)	&EW$_{6.7}$$+$EW$_{6.97}$~(eV) &Luminosity~(erg~s$^{-1})$\\
\hline
mCVs		&$\sim$120	&$\sim$260	&$\sim10^{32}$--$10^{34}$\\
nmCVs		&$\sim$70	&$\sim$530	&$\sim10^{30}$--$10^{32}$\\
ABs (\& CAs)	&$\le$20		&300--650		&$\sim10^{27}$--$10^{32}$\\
\hline
GCXE		&145$\pm$3 	&748$\pm$5	&\\
GBXE		&61$\pm$11	&603$\pm$20	&\\
GRXE		&97$\pm$12	&545$\pm$24	&\\
\hline
\end{tabular} 
\end{center}
Error is 1 $\sigma$ (68\% confidence) level. \\
\end{table*}

The SHs of stars depend on the mass (e.g., \cite{Hawkins1988,Gilmore2000,Bimmey2008}):
$\lesssim$100~pc for high-mass stars and $\gtrsim100$~pc for low-mass stars.  
Then the SH of CVs (mCVs+nmCVs) are in the range of 130--160~pc \citep{Patterson1984,Ak2008,Revnivtsev2008}.  
The spectral types of ABs are mostly G-K type with  small fraction ($\sim$15\%) of F type \citep{St1993}. 
Then the SH of ABs is $\sim$150--300~pc, similar to those of G--K type stars \citep{Gilmore2000}. 
The SH of CAs in the MS would be  $\sim$150--300~pc. However the CAs with
age of $\lesssim$10~Myrs, the CAs are not largely diffused out from the mother clouds, and hence the SH would be similar to MCs,  
$\lesssim$100~pc. 

\subsection{Galactic Bulge X-ray Emission (GBXE)} 

\citet{Revnivtsev2009} conducted a deep observation ($\sim$1~Msec) in the region of 
$(l, b)$=(\timeform{0.1D},\timeform{-1.4D}) (Chandra Bulge Field, CBF).  
Although the CBF is near the GCXE region, the  flux ratio of the iron K-shell lines of the GBXE and GCXE (GBXE/GCXE) are $\sim$10 
(see figure 1 left at $|b_{\ast}|\sim$\timeform{1.4D}).  Thus the CBF can be regarded as almost a pure GBXE region. 
In the CBF, \citet{Revnivtsev2009} and \citet{Hong2012} reported that $\sim$70--80\% flux (6.5--7.1 keV band) 
in the central region was resolved into point sources. 
However it is very surprising  that the profiles of the 6.5--7.1 keV flux as a function of 2--10 keV luminosity by  \citet{Hong2012} is
$\sim$2 times larger than that of  \citet{Revnivtsev2009} in the most important luminosity range of $10^{31}$--$10^{32}$~erg~s$^{-1}$.  
Furtheremore, about 20\% of the faintest point sources are unique in each point source lists. 
\citet{Hong2012} argued following his luminosity function that the major component is mCVs in contrast to major point source origin scenarios.  

\citet{Morihana2013}, on the other hand, reported that $\sim$50\% (2--8 keV band) of the full CBF field was resolved into point sources. 
In figure 13 of \citet{Morihana2013}, EW$_{6.7}$ of the CBF is $\sim$100~eV in the luminosity range of $\gtsim10^{32}$~erg~s$^{-1}$, 
where a  candidate source may be mCVs. It constantly increases in the range of $7\times10^{30}$--7$\times10^{31}$~erg~s$^{-1}$. 
This trend would be due to increasing contribution of nmCV and bright ABs,
which is against the argument of \citet{Hong2012}.
In the range of $\lesssim7\times10^{30}$~erg~s$^{-1}$, the EWs  become nearly constant at $\sim$300~eV, 
where main contributors would be faint ABs, CAs and others. 

We see many systematic errors and/or differences from author to author in the quantities of the point sources scenario even for the GBXE. 
These possible errors may be ignored in the luminosity range of $\gtrsim10^{30}$~erg~s$^{-1}$. 
Thus a robust conclusion may be that point sources occupy $\sim$50--70\% of the total GBXE flux 
in the $\gtrsim10^{30}$~erg~s$^{-1}$ range. 
The SH$_{6.7}$ and SH$_{6.97}$ of $\sim$310~pc and SH$_{6.4}$ of $\sim$160~pc are consistent with those of nmCVs and ABs. 
Also the EW$_{6.4}$ and EW$_{6.7}$$+$EW$_{6.97}$ are not inconsistent with the sum of nmCVs and ABs in any mixing ratio (table 3).
Thus we suspect that some fraction ($\sim$10--20\%) of the GBXE is mCVs, while  a major fraction ($\sim$40--50\%) are due to nmCVs and bright ABs, which covers mainly the luminosity range of 10$^{30}$--$10^{32}$~erg~s$^{-1}$. To explain another $\sim$ 30--50\%, more reliable information of the spectra of faint nmCVs, ABs, CAs or other objects is necessary.

\subsection {Galactic Ridge X-ray Emission (GRXE)} 

The SH$_{6.7}$  and  SH$_{6.97}$  are  $\sim$140~pc and  $\sim$100~pc, respectively. 
Within the error of $\sim$20--40 pc, these may be marginally consistent with  those of CVs and ABs. 
The EW$_{6.7}$$+$EW$_{6.97}$ of $\sim$550~eV is similar to the GBXE. 
Thus the origin of HP may be more or less similar to the GBXE: a large fraction is nmCVs$+$ABs in the luminosity range of 
$\lesssim10^{32}$~erg~s$^{-1}$.  
On the other hand, the EW$_{6.4}$ of $\sim$100~eV is $\sim$1.5--3 times larger
and SH$_{6.4}$ of $\sim70\pm{20}$~pc is smaller than any mixing ratio of mCVs, nmCvs and  ABs. 
We therefore more seriously examine the origin of the Fe\,\emissiontype{I}\,K$\alpha$ line than the case of the GBXE. 

Large excesses of the Fe\,\emissiontype{I}\,K$\alpha$ relative to the Fe\,\emissiontype{XXV}\,He$\alpha$ line
at $l$=\timeform {1.5D}--\timeform{3.5D} and
$l$=\timeform{330D}--\timeform{340D} (see figure 2) are also against the point source origin for the Fe\,\emissiontype{I}\,K$\alpha$ line. 
Since the excess is nearly 2 times of the average level, a significant fraction should be in unknown components, 
which have strong Fe\,\emissiontype{I}\,K$\alpha$ lines.
The SH$_{6.4}\sim70\pm{20}$~pc is similar to the MC distribution  \citep{Mathis2000,Stark2005}.
Therefore the Fe\,\emissiontype{I}\,K$\alpha$ line would mainly originate from  MCs. 

\citet{Molaro2014} claimed  that $\sim$10--30\% of the total luminosity of the GRXE would be the scattered flux of LMXBs.  
Using the best-fit parameters listed in table~2, the 5--8 keV band luminosity 
in the ($|l_*|= \timeform{10D}$--$\timeform{30D}$,  $|b_*|\le\timeform{0.5D}$ ) region is estimated to be $\sim6\times10^{36}$~erg~s$^{-1}$, 
while that of all the cataloged LMXBs in the same region is $\sim7\times10^{37}$~erg~s$^{-1}$ \citep{Liu2007}. 
The line-of-sight (on-plane) $N_{\rm H}$ from this region is $\sim4\times10^{22}$~cm$^{-2}$ (e.g., \cite{Ebisawa2005,Yasumi2014}).  
For simplicity, we assume a uniform density disk of 6 kpc radius and 70~pc thick around each  LMXB, then $N_{\rm H}$ 
averaging 4$\pi$ steradian around the LMXB is estimated to be $\sim4\times10^{21}$~cm$^{-2}$. 
Therefore the Thomson scattered luminosity is $\sim2\times10^{35}$~erg~s$^{-1}$.  
This is a few \% of the GRXE in this region, 
and hence we can safely conclude that the contribution of LMXBs to the Fe\,\emissiontype{I}\,K$\alpha$ flux of the GRXE is minor fraction. 

Another possible source of the Fe\,\emissiontype{I}\,K$\alpha$ line is ionization by low-energy cosmic-rays (LECR), 
either  protons (LECRp) or electrons (LECRe). 
\citet{Nobukawa2015} consistently explained that the Fe\,\emissiontype{I}\,K$\alpha$ excess at 
$l$=\timeform{1.5D}--\timeform{3.5D} is due to LECRp. 
In general, the most probable site of the LECRs is SNRs.  
However, our spectral data include no X-ray SNR. 
Also only a few SNRs are associated with the diffuse Fe\,\emissiontype{I}\,K$\alpha$ line \citep{Sato2014, Sato2015}. 

\subsection {Galactic Center X-ray Emission (GCXE)} 

Since the EWs of the GCXE are much larger than the GBXE and GRXE, the major origin of the GCXE cannot be the same 
as the GBXE and GRXE, namely nmCVs and ABs.
\citet{Uchiyama2011} found that the
Fe\emissiontype{XXV}\,He$\alpha$ line shows large excess over the stellar mass density model, 
assuming that all the GRXE and GBXE are due to integrated emission of point sources. 
The same excess over the real infrared star count profile is found by \citet{Yasui2015}.  

The smaller SHs of the GCXE (see table 2) than those of CVs and ABs also support
that GCXE needs large additional components with smaller SHs than CVs and ABs.
One plausible site of the GCXE  is the central molecular zone (CMZ) \citep{Tsuboi1999, Weinen2015}. 
Possible origin is a large amount of SNRs or very active star formation \citep{Koyama1986} 
in the CMZ.  
Another possibility is that the GCXE is due to the past high energy activities (flares of Sgr A$^*$), which is responsible for the XRNe, 
a recombining plasma \citep{Nakashima2013}, the Fermi bubble \citep{Su2010} and jet-like structures 
\citep {Koyama2003, Muno2008, Heard2013b}. 
All these possibilities can produce not only a HP responsible for the  Fe\,\emissiontype{XXV}\,He$\alpha$ and 
Fe\,\emissiontype{XXVI}\,Ly$\alpha$ lines but also non-thermal particles responsible for the Fe\,\emissiontype{I}\,K$\alpha$ line 
\citep{Nobukawa2015, Sato2015}. 
The excess of the Fe\,\emissiontype{I}\,K$\alpha$, 
Fe\,\emissiontype{XXV}\,He$\alpha$ and Fe\,\emissiontype{XXVI}\,Ly$\alpha$  lines
in the Sgr A East region \citep{Park2004, Koyama2007b} would be a good example. 

\bigskip
The authors are grateful to all members of the Suzaku team. 
This work was supported by 
the Japan Society for the Promotion of Science (JSPS) KAKENHI 
(No24540232, SY; No24740123, MN; No24540229, KK).
KKN is supported by Research Fellow of JSPS for Young Scientists.

\end{document}